\documentclass[aps,prd,10pt,notitlepage,nofootinbib,superscriptaddress,showkeys,showpacs]{revtex4-2}
\pdfoutput=1
\usepackage{amstext,amsmath,amssymb,amsfonts,bbm,euscript,color,dsfont}
\usepackage{epsfig}
\usepackage{hyperref}
\usepackage{amsthm}
\usepackage{color}
\usepackage{slashed}

\usepackage{latexsym}

\newcommand{\bea}{\begin{eqnarray}}	
\newcommand{\eea}{\end{eqnarray}}
\newcommand{\be}{\begin{equation}}	
\newcommand{\ee}{\end{equation}}
\newcommand{\beq}{\begin{equation}}	
\newcommand{\eeq}{\end{equation}}

\newcommand{\Z}{{\mathbb Z}}
\newcommand{\C}{{\mathbb C}}

\def\R{\relax\ifmmode {\mathbb R}  \else${\mathbb R}$\fi}
\def\C{\relax\ifmmode {\mathbb C}  \else${\mathbb C}$\fi}
\def\Z{\relax\ifmmode {\mathbb Z}  \else${\mathbb Z}$\fi}
\def\N{\relax\ifmmode {\mathbb N}  \else${\mathbb N}$\fi}
\def\I{\relax\ifmmode {\mathbb I}  \else${\mathbb I}$\fi}

\begin{document}

\title{The quantum electrodynamics for dyons} 


\author{D.O.R. Azevedo}\email{azevedo.dor@gmail.com}
 \affiliation{Instituto de F\'isica, Facultad de Ingenier\'ia, Universidad de la Rep\'ublica, J. H. y Reissig 565, 11000 Montevideo, Uruguay}
\author{O.M. Del Cima} \email{oswaldo.delcima@ufv.br}
\affiliation{Departamento de F\'isica, Universidade Federal de Vi\c cosa,
	Campus Universit\'ario, Avenida Peter Henry Rolfs s/n, 36570-900, Vi\c cosa, MG, Brazil}
 \affiliation{Ibitipoca Institute of Physics (IbitiPhys), Concei\c c\~ao do Ibitipoca, 36140-000, MG, Brazil}
\author{T.S. Dias} \email{thadeu.dias@ufv.br}
\affiliation{Departamento de F\'isica, Universidade Federal de Vi\c cosa,
	Campus Universit\'ario, Avenida Peter Henry Rolfs s/n, 36570-900, Vi\c cosa, MG, Brazil}
 \affiliation{Ibitipoca Institute of Physics (IbitiPhys), Concei\c c\~ao do Ibitipoca, 36140-000, MG, Brazil}
\author{E.D. Pereira} \email{emilio.drumond@ufv.br}
\affiliation{Departamento de F\'isica, Universidade Federal de Vi\c cosa,
	Campus Universit\'ario, Avenida Peter Henry Rolfs s/n, 36570-900, Vi\c cosa, MG, Brazil}
 \affiliation{Ibitipoca Institute of Physics (IbitiPhys), Concei\c c\~ao do Ibitipoca, 36140-000, MG, Brazil}



\begin{abstract}
The quantum electrodynamics for massive fermions carrying electric and magnetic charges, the dyons, is proposed based on the 1960's seminal works by Cabibbo, Ferrari, and Salam, with the gauge group being $U(1)\times U(1)$, which is associated to a vector field (photon) and a pseudo-vector field (metaphoton), wherein the Dirac quantization is set aside. At the tree level the spectrum consistency of the model is analyzed, and all continuous and discrete symmetries are established. The quantum analysis is performed by using the Becchi-Rouet-Stora (BRS) algebraic renormalization method, which is independent on any regularization scheme. Moreover, thanks to the presence of massless gauge fields, the Lowenstein-Zimmermann (LZ) subtraction scheme is required within the framework of the Bogoliubov-Parasiuk-Hepp-Zimmermann-Lowenstein (BPHZL) renormalization procedure. Finally, it is verified that the proposed model, the dyon quantum electrodynamics (dQED), is free from any anomaly and is multiplicative renormalizable at all orders in perturbation theory, proving, therefore, its quantum consistency.  
\end{abstract}

\maketitle
\section{Introduction}

Dirac's monopole, proposed in 1931~\cite{dirac1931}, served as a gateway to various models that sought to explain the existence of particles with intrinsic magnetic charges. These models include those by Cabibbo and Ferrari~\cite{cabibbo1962quantum}, Salam~\cite{salam1966magnetic}, Schwinger~\cite{Schwinger1968}, 't Hooft and Polyakov~\cite{thooftPolyakov}, and Nambu~\cite{nambu197476,Azevedodaniel}. Dirac's result showed that a magnetic charge cannot exist independently of an electric charge, as given by the Dirac quantization condition, $eg=n\hbar c/2$ ($n\in\mathbb{Z}$). This result arises from the singularity permitted by the 4-vector potential, specifically Dirac's strings, where each pole is situated at the end of the string. In the following years, Dirac revisited the concept of magnetic monopoles in 1948~\cite{dirac1948}. Subsequently, in 1962, Cabibbo and Ferrari proposed an alternative formulation of electromagnetism~\cite{cabibbo1962quantum}. This new formulation addressed the singularity problem encountered by Dirac introducing a second vector potential in electromagnetism, with the new potential associated with a magnetic charge, two gauge symmetries are given now, this aspect is particularly useful for investigating hidden or dark photons~\cite{deliyergiyev2016recent, bento2024classes, Bauer:2018onh} related to dark matter and physics beyond the Standard Model. The Cabibbo-Ferrari model also presents a reformulation of the $CPT$ theorem, as discussed by Ramsey~\cite{ramsey1958time}, in which discrete symmetries are expanded to include a new symmetry: the conjugation of the magnetic charge. On the other hand, in 1966, Salam suggested that magnetic monopoles could be associated to possible $C$-violation of electromagnetic interactions since Dirac quantization implying that the electric and magnetic charge are related. However, in order to avoid enormous vacuum polarization and $C$-violating effects in atomic physics~\cite{salam1966magnetic} -- owing to non-perturbative (high) value of magnetic charge -- the existence of a second ``photon'' could be used to circumvent the Dirac quantization, implying two distinct (massless) photons, one associated to electric charge conservation (photon) and the other associated to magnetic charge conservation (metaphoton). In 1968, Schwinger~\cite{Schwinger1968} idealized a model where the magnetic neutrality of matter resulted from the greater attractive force between two magnetic charges with opposite poles, in accordance with the Dirac quantization condition. In 1974, Nambu~\cite{nambu197476}, building on Dirac's work, studied quarks as carriers of magnetic charge connected by strings. Independently, 't Hooft and Polyakov~\cite{thooftPolyakov} developed models that demonstrated the possibility of creating magnetic monopoles within the framework of Grand Unified Theory (GUT). These GUT magnetic monopoles do not fit within the Standard Model and possess a mass limit much greater than any particle produced in laboratories~\cite{rajantie2016magnetic}.

Notwithstanding the fact that models incorporating magnetic charges are very promising, they face a significant issue that challenges the validity of perturbation theory. Specifically, expansions in terms of electric and magnetic charges may diverge. While a viable approximation might exist based on the Dirac quantization condition, the superstrong coupling between photons and magnetic monopoles, as discussed in Rabl~\cite{Rabl:1969gx}, lies beyond the range of perturbative methods. Moreover, Datta~\cite{Datta:1983um} shows that the small value of the electric fine-structure constant that is essential for the stability and binding of electronic systems can influence magnetic monopoles in a way that may lead to their collapse. However, the Cabibbo-Ferrari formulation that eliminates Dirac's string is not necessarily constrained by the Dirac quantization condition. In 1966, Salam~\cite{salam1966magnetic} showed that the magnetic charge does not need to be related to the electric charge. Similarly, in 1996, Singleton~\cite{singleton1996does} argued that a large non-perturbative magnetic coupling would lead to a phase transition in which magnetic charge becomes confined and the photon acquires a mass. The apparent lack of experimental evidence for a massive photon can thus be used to argue against the existence of such a large, non-perturbative, Dirac-type Abelian magnetic charge. Govaerts~\cite{Govaerts:2023iqf} has described an extended electromagnetic theory that is also not restricted by the Dirac quantization condition. Although the studies by Rabl, Datta, Singleton, and Govaerts do not constitute a definitive proof of the non-existence of Dirac-type magnetic monopoles, they nonetheless provide a motivating context for the development of our model, the dyon quantum electrodynamics (dQED)~\cite{msc-thadeu}. Recently, the Cabibbo-Ferrari model was studied by adopting the Gupta-Bleuler procedure in the Feynman-'t Hooft gauge~\cite{anwar}, and \cite{vento} proposed that the Cabibbo-Ferrari model could be related to the dark sector. Most recently, the non-linear formulation of the dQED model as a one-loop Euler-Heisenberg effective theory is presented making use of the Schwinger's proper-time formalism~\cite{azevedo}. The search for magnetic monopoles, which involves experiments with ATLAS/MoEDAL (Monopole and Exotics Detector at the LHC), includes Dirac magnetic monopoles, high electric charge objects (HECOs), up-down quark matter, and Q balls. These are collectively referred to as highly charged particles and ionizing agents (HIPs)~\cite{ATLAS:2023esy}.

Our goal is to build up a new model that incorporates not only the formalism of Cabibbo and Ferrari, but also the Salam's proposal. Our basic hypothesis involves extending the gauge group from $U(1)$ to $U(1)\times U(1)$, thereby introducing a new potential associated with the magnetic charge. Furthermore, the Dirac quantization condition is not taken into consideration -- as suggested by Salam~\cite{salam1966magnetic} so as to prevent measureless vacuum polarization and $C$-violating effects in atomic physics -- allowing, as a byproduct, the description of magnetic monopoles within the framework of perturbation theory. The fermionic matter field is taken to be a massive spinor that carries both electric and magnetic charges, in other words they are dyons. This framework defines here what we refer to as the dyon quantum electrodynamics (dQED).

In this work, we propose and investigate, the dyon quantum electrodynamics (dQED), a Cabibbo-Ferrari-Salam based model, focusing on its tree level and quantum level consistency. The paper is organized as follows, in Section II, we introduce the dQED action, along with the gauge symmetries, the Becchi-Rouet-Stora (BRS) transformations, and the discrete symmetries. In Section III, we compute the tree-level propagators and analyze the spectral consistency of the model. Section IV presents the Slavnov-Taylor identity and other tree-level symmetries satisfied by the action. Finally, in Section V, at all orders in perturbation theory, we explore potential breakings of the Slavnov-Taylor identity, namely, searching for a gauge anomaly, furthermore we discuss the multiplicative renormalizability of the dQED model.

\section{The dyon quantum eletrodynamics (dQED) model and its symmetries}

Starting from Maxwell's equations, we can modify them by introducing a magnetic source, {\it i.e.}, a magnetic monopole. The new set of equations can now be written as follows:\footnote{The dual tensor $\tilde{F}^{\mu\nu}$ is defined as 
$\tilde{F}^{\mu\nu}=\frac{1}{2}\varepsilon^{\mu\nu\rho\sigma}F_{\mu\nu}$, where the covariant and contravariant forms of the completely antisymmetric Levi-Civita tensor in four dimensions are related by $\varepsilon^{\mu\nu\rho\sigma}=-\varepsilon_{\mu\nu\rho\sigma}$. This relationship arises because the metric is given by $\eta_{\mu\nu}=\text{diag}(1,-1,-1,-1)$.},
\begin{equation}
    \begin{matrix}
        \partial_{\mu}F^{\mu\nu}=j^{\nu}~,\\
        \partial_{\mu}\tilde{F}^{\mu\nu}=g^{\nu}~,
    \end{matrix}
\end{equation}
where $j^{\nu}$ is the electric current while $g^{\nu}$ is the magnetic current. Cabibbo and Ferrari addressed this problem by modifying the field strength $F_{\mu\nu}$, to the two-potential approach, resulting the Cabibbo-Ferrari tensor~\cite{cabibbo1962quantum} given by:
\begin{equation}\label{eq10}
\begin{matrix}
\mathcal{F}^{\mu\nu}=\partial^{\mu}A^{\nu}-\partial^{\nu}A^{\mu}+\varepsilon^{\mu\nu\rho\sigma}\partial_{\rho}B_{\sigma}~,\\
\mathcal{F}_{\mu\nu}=\partial_{\mu}A_{\nu}-\partial_{\nu}A_{\mu}-\varepsilon_{\mu\nu\rho\sigma}\partial^{\rho}B^{\sigma}~, 
\end{matrix}
\end{equation}
with $A_{\mu}$ being the usual four-vector (photon) associated to electric charge conservation, whereas $B_{\mu}$ being the new four-vector (metaphoton) related to the magnetic charge conservation. The introduction of a second potential helps to avoid the singularity problem. Additionally, $A_{\mu}$ transforms like a four-vector, whereas $B_{\mu}$ transforms like a pseudo four-vector (or ``axial vector''), which results from the Levi-Civita tensor, thus we have two sets of vector fields, $A_\mu$ and $B_\mu$, which describe electric and magnetic charges, respectively. This setup introduces extra degrees of freedom, and the Cabibbo-Ferrari consequently has a $U(1)\times U(1)$ gauge symmetry, extending the usual electromagnetic gauge symmetry.

The classical action for the dyon quantum electrodynamics (dQED)~\cite{msc-thadeu}, based on the Cabibbo-Ferrari model~\cite{cabibbo1962quantum}, reads:
\begin{equation}\label{eq11}
\Sigma_{\rm dQED}=\int d^{4}x \left[-\frac{1}{4}\mathcal{F}_{\mu\nu}\mathcal{F}^{\mu\nu}+\bar{\psi}(\imath \gamma^{\mu}D_{\mu}-m )\psi \right]~,
\end{equation}
where $\psi$ is the Dirac spinor associated to a dyon carrying electric charge ($e$) and magnetic charge ($g$), with mass ($m$), $\mathcal{F}_{\mu\nu}$ is the Cabibbo-Ferrari field strength (\ref{eq10}), and the covariant derivative $D_{\mu}\psi=(\partial_{\mu}-\imath eA_{\mu}-\imath gB_{\mu})\psi$. Moreover, it is straightforward to verify that the dQED action (\ref{eq11}) is invariant under the gauge transformations, $\delta_{\theta}$ and $\delta_{\chi}$, displayed below:
\begin{equation}\label{eq18}
    \begin{matrix}
    \displaystyle\delta_{\theta}A_{\mu}=-\partial_{\mu}\theta~, & \delta_{\theta}B_{\mu}=0~, & \delta_{\theta}\psi=-\imath e\theta\psi~, & \delta_{\theta}\bar{\psi}=\imath e\theta\bar{\psi}~,\\
    \displaystyle\delta_{\chi}A_{\mu}=0~, & \delta_{\chi}B_{\mu}=-\partial_{\mu}\chi~, & \delta_{\chi}\psi=-\imath g\chi\psi~, & \delta_{\chi}\bar{\psi}=\imath g\chi\bar{\psi}~,
    \end{matrix}
\end{equation}
where $\theta$ and $\chi$ are infinitesimal local parameters.

Let us rewrite the action (\ref{eq11}) explicitly in terms of the potential vectors, $A_{\mu}$ and $B_{\mu}$, by using the definitions (\ref{eq10}) together with the identity of the $\varepsilon^{\mu\nu\rho\sigma}$ tensor in four space-time dimensions \footnote{$\varepsilon^{\mu\nu\rho\sigma}\varepsilon_{\mu\nu\lambda\delta}=-2(\delta^{\rho}_{\lambda}\delta^{\sigma}_{\delta}-\delta^{\sigma}_{\lambda}\delta^{\rho}_{\delta})$.}, as a result the dQED action (\ref{eq11}) can now be expressed as follows:
\begin{equation}\label{eq15}
    \Sigma_{\rm dQED}=\int d^{4}x \left[-\frac{1}{4}F_{\mu\nu}F^{\mu\nu}-\frac{1}{4}G_{\mu\nu}G^{\mu\nu}+\bar{\psi}(\imath \gamma^{\mu}D_{\mu}-m )\psi \right]~,
\end{equation}
such that,
\begin{equation}
   F^{\mu\nu}=\partial^{\mu}A^{\nu}-\partial^{\nu}A^{\mu}~,~~G^{\mu\nu}=\partial^{\mu}B^{\nu}-\partial^{\nu}B^{\mu}~,~~D_{\mu}=\partial_{\mu}-\imath eA_{\mu}-\imath gB_{\mu}~,
\end{equation}
where $e$ is the electric charge, $g$ is the magnetic charge, and $m$ is the mass parameter of the dyon fermion field.

Through the use of the Becchi-Rouet-Stora (BRS) method~\cite{becchi,piguet-rouet}, the gauge-fixing action is build up as: 
\begin{equation}\label{sgf}
	\begin{split}
		\Sigma_{\rm gf} &= \mathfrak{s} \int d^{4}x \left[ \bar{c}\partial^{\mu}A_{\mu}+\frac{\alpha}{2} \bar{c}b+\bar{\xi}\partial^{\mu}B_{\mu}+\frac{\beta}{2} \bar{\xi}\pi \right]\\
		&=\int d^{4}x \left[ b\partial^{\mu}A_{\mu}+\bar{c}\Box c+\frac{\alpha}{2} b^{2}+\pi\partial^{\mu}B_{\mu}+\bar{\xi}\Box \xi+\frac{\beta}{2} \pi^{2} \right]~,
	\end{split}    
\end{equation}
with $\alpha$ and $\beta$ being the gauge-fixing parameters. Additionally to that, the BRS transformations~\cite{becchi,piguet-rouet} are given by:
\begin{equation}\label{eq29}
     \begin{matrix}
\mathfrak{s}A_{\mu}=-\partial_{\mu}c~, & \mathfrak{s}B_{\mu}=-\partial_{\mu}\xi~, \\
\mathfrak{s}\psi=-\imath(ec+g\xi)\psi~, & \mathfrak{s}\bar{\psi}=\imath(ec+g\xi)\bar{\psi}~, \\
\mathfrak{s}c=0~, & \mathfrak{s}\xi=0~, \\
\mathfrak{s}\bar{c}=b~, & \mathfrak{s}\bar{\xi}=\pi~, \\
\mathfrak{s}b=0~, & \mathfrak{s}\pi=0~,
\\
\end{matrix}
\end{equation}
where ($c$,$\xi$) are the Faddeev-Popov ghosts~\cite{faddeev1967feynman} with Faddeev-Popov charge (ghost number) $+1$, the antighosts ($\bar{c}$,$\bar{\xi}$) have ghost number $-1$ and the Nakanishi-Lautrup~\cite{lautrup-nakanishi} fields ($b$,$\pi$) have ghost number $0$, beyond that, the BRS operator $\mathfrak{s}$ is nilpotent, {\it i.e.} $\mathfrak{s}^2=0$. 

Taking into consideration the quantization that will follow, since the BRS transformations of the fermionic field are non-linear, we have to include an action in which these non-linear BRS transformations are coupled to BRS invariant external sources (antifields), so as to control their renormalization at quantum level:
\begin{equation}\label{eq34}
    \Sigma_{\rm ext}=\int d^{4}x \left[ \bar{Y}\mathfrak{s}\psi - \mathfrak{s}\bar{\psi}Y \right]~.
\end{equation}
Owing to the presence of massless gauge fields, namely $A_{\mu}$ and $B_{\mu}$, the Lowenstein-Zimmermann (LZ) subtraction scheme in the context of the Bogoliubov-Parasiuk-Hepp-Zimmermann-Lowenstein (BPHZL) renormalization method~\cite{Low} is adopted in order to deal with the infrared divergences that shall arise in the ultraviolet subtraction process. Accordingly, the Lowenstein-Zimmermann mass term action for the gauge fields is written as,
\begin{equation}\label{IR}
\Sigma_{\rm IR}=\int d^{4}x\left[\frac{1}{2}M_{A}^{2}(s-1)A_{\mu}A^{\mu}+\frac{1}{2}M_{B}^{2}(s-1)B_{\mu}B^{\mu} \right]~,
\end{equation}
where the LZ mass terms, $M_{A}^{2}(s-1)$ and $M_{B}^{2}(s-1)$, allow the use of a momentum-space subtraction scheme without introducing spurious infrared singularities. The Lowenstein-Zimmermann parameter $s$ lies in the range $0 \leq s\leq 1$, and plays the role of an additional subtraction variable (like the external momentum) in the BPHZL renormalization program so that the theory describing truly massless particles is recovered for $s=1$~\cite{del2016symanzik}. Here, it is worth mentioning that, despite the Lowenstein-Zimmermann mass terms for the gauge fields be not gauge invariant at the tree level, the gauge invariance is preserved at the quantum level, this is a peculiarity of the Abelian case~\cite{piguet-rouet}, this was discussed in detail for the massive vector-meson $U(1)$ QED~\cite{low-schroer} through the BPHZ renormalization scheme. Moreover, although they are massless, the Faddeev-Popov ghosts are free fields, they decouple, so no Lowenstein-Zimmermann mass term needs to be introduced for them.

Finally, the tree-level action for the dyon quantum electrodynamics (dQED) is defined as:
\begin{equation}\label{eq35}
    \Gamma^{(0)}\equiv\Gamma^{(0)}_{(s-1)}=\Sigma_{\rm dQED}+\Sigma_{\rm IR}+\Sigma_{\rm gf}+\Sigma_{\rm ext}~.
\end{equation}
Thanks to the presence of the Lowenstein-Zimmermann mass term, we see that the BRS symmetry of $\Gamma^{(0)}$ (\ref{eq35}) is broken:
\begin{equation}
\mathfrak{s}\Gamma^{(0)}=(s-1)\Delta_{\rm lin}~,
\end{equation}
with $\Delta_{\rm lin}$ being a linear breaking in the quantum fields:
\begin{equation}
\Delta_{\rm lin}=\int d^{4}x\left[-M_{A}^{2}A^{\mu}\partial_{\mu}c-M_{B}^{2}B^{\mu}\partial_{\mu}\xi \right]~,
\end{equation}
since the ghosts $c$ and $\xi$ are free, they decouple, therefore the breaking $\Delta_{\rm lin}$ gets no radiative corrections. Also, it should be pointed out that the BRS invariance is restored by setting the Lowenstein-Zimmermann parameter $s=1$, that is, $\mathfrak{s}\Gamma^{(0)}|_{s=1}=0$.

Regarding the discrete symmetries, namely parity, time reversal and charge conjugation, particularly the charge conjugation, since we are dealing with dyons, particular care is required. As suggested by Ramsey, based on Dirac's work, it was concluded that in a theory including the effects of magnetic charges, the usual $CPT$ theorem~\cite{Luders,Greenberg} should be modified to an $MEPT$ theorem~\cite{ramsey1958time}, where $M$ represents magnetic charge conjugation, $E$ electric charge conjugation, $T$ time reversal, and $P$ parity.

The $MEPT$ theorem can be expressed in different ways. Here, due to the presence of a dyon, which carries magnetic and electric charges, we choose the prescription, ${\widetilde C}PT$, where ${\widetilde C} = ME$ encloses both types of charge conjugation: the electric charge conjugation ($E$) and the magnetic charge conjugation ($M$). Furthermore, the magnetic charge $g$ is assumed a pseudoscalar under parity ($P$) and time reversal ($T$) transformations, while the electric charge $e$ is taken as a scalar under the same transformations. Accordingly, by adopting the Dirac representation of the $\gamma$-matrices~\cite{itzykson}, the discrete transformations read as follows:

\noindent \underline{{\it Charge Conjugation} (${\widetilde C}$)}:
\begin{align}
&\psi\overset{{\widetilde C}}{\longrightarrow}\psi^{{\widetilde C}}={\widetilde C}\bar{\psi}^{T}~,~~~~\bar{\psi}\overset{{\widetilde C}}{\longrightarrow}\bar{\psi}^{{\widetilde C}}=-\psi^{T}{\widetilde C}^{-1}~,\nonumber\\
&Y\overset{{\widetilde C}}{\longrightarrow}{\widetilde C}\bar{Y}^{T}~,~~~~\bar{Y}\overset{{\widetilde C}}{\longrightarrow}-Y^{T}{\widetilde C}^{-1}~,\nonumber\\
&A_{\mu}\overset{{\widetilde C}}{\longrightarrow}-A_{\mu}~,~~~~B^{\mu}\overset{{\widetilde C}}{\longrightarrow}-B_{\mu}~,\nonumber\\
&\varphi\overset{{\widetilde C}}{\longrightarrow} -\varphi~,~~~\varphi=\{c,\bar{c},b\}~,\nonumber\\
&\Xi\overset{{\widetilde C}}{\longrightarrow} -\Xi~,~~~\Xi=\{\xi,\bar{\xi},\pi\}~,\nonumber\\
&{\widetilde C}(\gamma^{\mu})^{T}{\widetilde C}^{-1}=-\gamma^{\mu}~,~~~~{\widetilde C}(\gamma^{5})^{T}{\widetilde C}^{-1}=\gamma^{5}~,\label{charge}
\end{align}
with ${\widetilde C}$ being the total charge conjugation matrix, \textit{i.e.}, it includes both electric charge conjugation ($E$) and magnetic charge conjugation ($M$), obeying ${\widetilde C}^{2}=1$.

\noindent \underline{{\it Parity} ($P$)}:
\begin{align}
&x=\overset{P}{\longrightarrow}(x^{0}~,-\Vec{x})~, \nonumber\\
&\psi\overset{P}{\longrightarrow}\psi^{P}=\gamma^{0}\psi~,~~~~\bar{\psi}\overset{P}{\longrightarrow}\bar{\psi}^{P}=\bar{\psi}\gamma^{0}~,\nonumber\\
&Y\overset{P}{\longrightarrow}Y^{P}=\gamma^{0}Y~,~~~~\bar{Y}\overset{P}{\rightarrow}\bar{Y}^{P}=\bar{Y}\gamma^{0}~,\nonumber\\
&A_{\mu}\overset{P}{\longrightarrow}A^{\mu}~,~~~~B_{\mu}\overset{P}{\longrightarrow}-B^{\mu}~,\nonumber\\
&\varphi\overset{P}{\longrightarrow}\varphi~,~~~\varphi=\{c,\bar{c},b\}~, \nonumber\\
&\Xi\overset{P}{\longrightarrow} -\Xi~,~~~\Xi=\{\xi,\bar{\xi},\pi\}~,\nonumber\\
&e\overset{P}{\longrightarrow}e~,~~~~g\overset{P}{\longrightarrow}-g~,\label{parity}
\end{align}
with $P$ being the usual parity matrix, obeying $P^{2}=1$.

\noindent \underline{{\it Time Reversal} ($T$)}:
\begin{align}
&x\overset{T}{\longrightarrow}(-x^{0},\Vec{x})~,\nonumber\\
&\psi\overset{T}{\longrightarrow}\psi^{T}=[\gamma^{1}\gamma^{3}]\psi~,~~~~\bar{\psi}\overset{T}{\longrightarrow}\bar{\psi}^{T}=\bar{\psi}[-\gamma^{1}\gamma^{3}]~,\nonumber\\
&Y\overset{T}{\longrightarrow}Y^{T}=[\gamma^{1}\gamma^{3}]Y~,~~~~\bar{Y}\overset{T}{\longrightarrow}\bar{Y}^{T}=\bar{Y}[-\gamma^{1}\gamma^{3}]~,\nonumber\\
&A_{\mu}\overset{T}{\longrightarrow}A^{\mu}~,~~~~B_{\mu}\overset{T}{\longrightarrow}-B^{\mu}~,\nonumber\\
&\varphi\overset{T}{\longrightarrow} -\varphi~,~~~\varphi=\{c,\bar{c},b\}~,\nonumber\\
&\Xi\overset{T}{\longrightarrow} \Xi~,~~~\Xi=\{\xi,\bar{\xi},\pi\}~,\nonumber\\
&e\overset{T}{\longrightarrow}e~,~~~~g\overset{T}{\longrightarrow}-g~,\label{time}
\end{align}
here $T$ is the usual time reversal matrix being antiunitary, obeying $T^{2}=1$.

Now, it can be checked that the action $\Gamma^{(0)}$ (\ref{eq35}) is invariant under ${\widetilde C}$, $P$, $T$ and ${\widetilde C}PT$. Finally, we may also note that the tree-level action (\ref{eq35}) is invariant under the following discrete exchange -- a duality -- transformation of the fields and parameters:
\begin{align}
&\{A_{\mu},B_{\mu}\} \longrightarrow \{B_{\mu},A_{\mu}\} ~,~~ \{e,g\} \longrightarrow \{g,e\} ~,~~ \{M_{A},M_{B}\} \longrightarrow \{M_{B},M_{A}\} ~,~~ \{\alpha,\beta\} \longrightarrow \{\beta,\alpha\}~, \nonumber\\
& \{c,\bar{c},b\}\longrightarrow \{\xi,\bar{\xi},\pi\} ~,~~ \{\xi,\bar{\xi},\pi\}\longrightarrow \{c,\bar{c},b\} ~,~~
\Gamma^{(0)}\longrightarrow\Gamma^{(0)}~. \label{trocatrocafields}
\end{align}

\section{Spectral analysis}

By switching off the coupling constants ($e=g=0$), we get the free part of the action, $\Gamma^{(0)}_{\rm free}=\Gamma^{(0)}|_{e=g=0}$, enabling the computation of the propagators for all the fields, thus they read:
\begin{align}\label{eq49}
&\displaystyle\Delta^{AA}_{\mu\nu}(k,s)=-\imath \left[ \frac{1}{k^{2}-M_{A}^{2}(s-1)^{2}}\left( \eta_{\mu\nu}-\frac{k_{\mu}k_{\nu}}{k^{2}} \right)+\frac{\alpha}{k^{2}-\alpha M^{2}_{A}(s-1)^{2}}\left( \frac{k_{\mu}k_{\nu}}{k^{2}} \right) \right]~, \nonumber\\
&\displaystyle\Delta^{BB}_{\mu\nu}(k,s)=-\imath \left[ \frac{1}{k^{2}-M_{B}^{2}(s-1)^{2}}\left( \eta_{\mu\nu}-\frac{k_{\mu}k_{\nu}}{k^{2}} \right)+\frac{\beta}{k^{2}-\beta M^{2}_{B}(s-1)^{2}}\left( \frac{k_{\mu}k_{\nu}}{k^{2}} \right) \right]~,\nonumber\\
&\displaystyle\Delta^{Ab}_{\mu}(k)=\Delta^{B\pi}_{\mu}(k)=-\frac{k_{\mu}}{k^{2}}~,\nonumber\\
&\displaystyle\Delta^{bb}(k)=\Delta^{\pi\pi}(k)=0~,\nonumber\\
&\displaystyle\Delta^{c\bar{c}}(k)=\Delta^{\xi\bar{\xi}}(k)=-\frac{\imath}{k^{2}}~,\nonumber\\
&\displaystyle\Delta^{\psi\bar{\psi}}(k)=\imath\frac{(\gamma^{\mu}k_{\mu}+m)}{k^{2}-m^{2}}~.
\end{align}

In Table \ref{tab1} we collect the ultraviolet (UV) dimensions, infrared (IR) dimensions and ghost numbers ($\Phi\Pi$). To obtain the UV dimension ($d$) and IR dimension ($r$) of any fields, $X$ and $Y$, use has been made of the UV and IR asymptotic behavior of their propagator, $\Delta_{XY}(k,s)$, where $d_{XY}$ is the UV asymptotic power for $(k,s)\longrightarrow\infty$, while $r_{XY}$ gives the IR asymptotic power for $(k,s-1)\longrightarrow 0$. The UV ($d$) and IR ($r$) dimensions of the fields, $X$ and $Y$, are chosen to satisfy the following inequalities\footnote{The derivative $\partial_{\mu}$ has UV and IR dimensions given by $d=r=1$.}~\cite{piguet2}:
\begin{equation}
d_{X} + d_{Y} \geq  4 + d_{XY}~~~\text{and}~~~r_{X} + r_{Y}\leq 4 + r_{XY} ~.
\end{equation}

\begin{table}[h]
	\begin{tabular}{| c | c | c | c | c | c | c | c | c | c | c | c | c | c |}
		\hline
	 & $A_{\mu}$ & $B_{\mu}$ & $\psi$ & $c$ & $\bar{c}$ & $b$ & $\xi$ & $\bar{\xi}$ & $\pi$ & $Y$ & $s-1$ & $s$\\
		\hline
		$d$ & $1$ & $1$ & $3/2$ & $0$ & $2$ & $2$ & $0$ & $2$ & $2$ & $5/2$ & 1 & 1\\
            \hline
    $r$ & $1$ & $1$ & $2$ & $0$ & $2$ & $2$ & $0$ & $2$ & $2$ & $2$ & 1 & 0\\
		\hline
		$\Phi\Pi$ & $0$ & $0$ & $0$ & $1$ & $-1$ & $0$ & $1$ & $-1$ & $0$ & $-1$ & 0 & 0\\
		\hline
	\end{tabular}
	\caption{UV dimension ($d$), IR dimension ($r$) and ghost number ($\Phi\Pi$).}
	\label{tab1}
\end{table}

\subsection{Unitarity and causality}

In order to avoid ghosts (negative-norm states) and tachyons (superluminal states), unitarity and causality have to be verified. This is done through coupling the propagators to external currents $\{\mathcal{J}_{\Phi_{i}} \}$ compatible with symmetries of the model, by computing the residues of the current-current amplitudes ($\mathcal{A}_{\Phi_{i}\Phi_{j}}$) at the poles (physical states):
\begin{equation}\label{eq50}
    \mathcal{A}_{\Phi_{i}\Phi_{j}}=\mathcal{J}_{\Phi_{i}}^{*}(k)\left< \Phi_{i}(k)\Phi_{j}(k)\right>\mathcal{J}_{\Phi_{j}}(k)~,
\end{equation}
checking if the imaginary part of the residues are positive definite at the poles -- $\Im\{\text{Res}\mathcal{A}_{\Phi_{i}\Phi_{j}}|_\text{poles}\}>0$ -- and also if the poles are non-negative, subluminal ($k^2<0$) or luminal ($k^2=0$) states. Consequently, if is such the case, the $S$-matrix is unitarity, allowing therefore the counting of the propagating physical degrees of freedom described by the field $\Phi_{i}$~\cite{Wadojackiwpi}.

First, let us analyze the case of the propagators of the vector fields, $A_{\mu}$ and $B_{\mu}$, displayed in (\ref{eq49}), by taking $s=1$, thus $\Delta^{AA}_{\mu\nu}(k,s)|_{s=1}\equiv\Delta^{AA}_{\mu\nu}(k)$ and $\Delta^{BB}_{\mu\nu}(k,s)|_{s=1}\equiv\Delta^{BB}_{\mu\nu}(k)$. The vector currents, $\mathcal{J}^{\mu}_{A}$ and $\mathcal{J}^{\mu}_{B}$, may be expanded in terms of a four-dimensional complete basis $\{ k^{\mu},\Tilde{k}^{\mu},\epsilon^{\mu},\Tilde{\epsilon}^{\mu} \}$ in the momentum space as follows:
\begin{align}\label{eqscurrent}
&\mathcal{J}^{\mu}_{A}=\lambda_{1}k^{\mu}+\lambda_{2}\Tilde{k}^{\mu}+\lambda_{3}\epsilon^{\mu}+\lambda_{4}\Tilde{\epsilon}^{\mu}~,\nonumber \\
&\mathcal{J}^{\mu}_{B}=\chi_{1}k^{\mu}+\chi_{2}\Tilde{k}^{\mu}+\chi_{3}\epsilon^{\mu}+\chi_{4}\Tilde{\epsilon}^{\mu}~.\nonumber
\end{align}
However, since we are dealing with a massless pole, where the dispersion relation is $k^{\mu}k_{\mu}=0$, the four-vector basis can be chosen as $k^{\mu}=(\omega,0,0,\omega)$, $\tilde{k}^{\mu}=(\omega,0,0,-\omega)$, $\epsilon^{\mu}=(0,\epsilon,\varepsilon,0)$ and $\tilde{\epsilon}^{\mu}=( 0,\varepsilon,-\epsilon,0)$, where $\omega$, $\epsilon$ and $\varepsilon$ are free real parameters. In addition to, the vector currents, $\mathcal{J}^{\mu}_{A}$ and $\mathcal{J}^{\mu}_{B}$, satisfy the following conservation conditions:
\begin{equation}
    k_{\mu}\mathcal{J}^{\mu}_{A}=0~\text{and}~k_{\mu}\mathcal{J}^{\mu}_{B}=0~, 
\end{equation}
as a result, the currents read as below:
\begin{align}\label{eq51}
&\displaystyle\mathcal{J}^{\mu}_{A}=(\lambda_{1}\omega,\lambda_{3}\epsilon+\lambda_{4}\varepsilon,\lambda_{3}\varepsilon-\lambda_{4}\epsilon,\lambda_{1}\omega)~,\nonumber \\
&\displaystyle\mathcal{J}^{\mu}_{B}=(\chi_{1}\omega,\chi_{3}\epsilon+\chi_{4}\varepsilon,\chi_{3}\varepsilon-\chi_{4}\epsilon,\chi_{1}\omega)~. 
\end{align}
The current-current amplitudes for the vector fields, $A^{\mu}$ and $B^{\mu}$, are finally given by:
\begin{align}\label{eq52}
&\mathcal{A}_{AA}=i\frac{1}{k^{2}}(|\lambda_{3}|^{2}+|\lambda_{4}|^{2})~,\nonumber\\
&\mathcal{A}_{BB}=i\frac{1}{k^{2}}(|\chi_{3}|^{2}+|\chi_{4}|^{2})~.
\end{align}
In sequence, for the propagators (\ref{eq49}) involving the fields $b$, $\pi$, $c$, $\bar{c}$, $\xi$ and $\bar{\xi}$, the current-current amplitudes read:
\begin{align}\label{eq53}
&\mathcal{A}_{Ab}=\mathcal{A}_{B\pi}=\mathcal{A}_{bb}=\mathcal{A}_{\pi\pi}=0~,\nonumber\\
&\mathcal{A}_{c\bar{c}}=-i\frac{1}{k^{2}}\mathcal{J}_{c}^{*}(k)\mathcal{J}_{\bar{c}}(k)~,\nonumber\\
&\mathcal{A}_{\xi\bar{\xi}}=-i\frac{1}{k^{2}}\mathcal{J}_{\xi}^{*}(k)\mathcal{J}_{\bar{\xi}}(k)~.
\end{align}
Lastly, the current-current amplitude in the case of the fermion field, $\psi$, since the pole is massive, with $k^{2}=m^{2}$, the momentum can be chosen in the rest frame, $k^{\mu}=(m,0,0,0)$, therefore it follows that:
\begin{align}\label{eq62}
\mathcal{J}_{\psi}=\begin{bmatrix}
\sigma_{1} \\
\sigma_{2} \\
\sigma_{3} \\
\sigma_{4}
\end{bmatrix}~~\text{and}~~
\mathcal{\bar{J}}_{\psi}=\begin{bmatrix}
\sigma_{1}^{*} & \sigma_{2}^{*} & -\sigma_{3}^{*} & -\sigma_{4}^{*} \\
\end{bmatrix}~~\text{with}~~\mathcal{A}_{\psi\bar{\psi}}=i\frac{2m}{k^{2}-m^{2}}\left( |\sigma_{1}|^{2}+|\sigma_{2}|^{2} \right).
\end{align}

Accordingly, collecting all the amplitudes computed, (\ref{eq52}), (\ref{eq53}) and (\ref{eq62}), and taking their imaginary part of the residue at their poles, we get:
\begin{align}\label{eqimaginary}
&\Im\{\text{Res}\mathcal{A}_{AA}|_{k^{2}=0}\}>0 ~,~~ \Im\{\text{Res}\mathcal{A}_{BB}|_{k^{2}=0}\}>0~, \nonumber\\
&\Im\{\text{Res}\mathcal{A}_{Ab}|_{k^{2}=0}\}=\Im\{\text{Res}\mathcal{A}_{bb}|_{k^{2}=0}\}=\Im\{\text{Res}\mathcal{A}_{B\pi}|_{k^{2}=0}\}=\Im\{\text{Res}\mathcal{A}_{\pi\pi}|_{k^{2}=0}\}=0~,\nonumber\\
&\Im\{\text{Res}\mathcal{A}_{c\bar{c}}|_{k^{2}=0}\}<0 ~,~~ 
\Im\{\text{Res}\mathcal{A}_{\xi\bar{\xi}}|_{k^{2}=0}\}<0~,\nonumber\\
&\Im\{\text{Res}\mathcal{A}_{\psi\bar{\psi}}|_{k^{2}=m^{2}}\}>0~.
\end{align}

From the final results presented in (\ref{eqimaginary}), we conclude that the vector fields, $A^{\mu}$ and $B^{\mu}$, propagate two massless degrees of freedom each, while there are no massless modes propagating in the Nakanishi-Lautrup field sector. However, the both Faddeev-Popov pair ghost-antighost, ($c$,$\bar{c}$) and ($\xi$,$\bar{\xi}$), propagate two massless degrees of freedom each which take care of the two spurious degrees of freedom arising from the longitudinal sector of the both vector fields. Moreover, the dyon fermionic field ($\psi$) carries four massive degrees of freedom. In summary, we conclude that the tree-level dyon quantum electrodynamics (dQED) is free from superluminal states (tachyons) and fulfills the necessary conditions for $S$-matrix unitarity, it is free from negative-norm states (ghosts). Beyond that, owing to the ultraviolet behaviour of the vector fields and the fermion field propagators, as well as the structure of their interaction vertices, the Froissart-Martin bound~\cite{itzykson,chaichian} shall not be violated, so the interactions might not spoil unitarity either.

\section{Slavnov-Taylor and another identities}

With the aim to represent the symmetries in a functional way, we begin with the Slavnov-Taylor operator ($\mathcal{S}$) acting on an arbitrary functional ($\mathcal{F}$):
\begin{equation}\label{ST}
    \mathcal{S}(\mathcal{F})=\int d^{4}x~\left[ -\partial_{\mu}c\frac{\delta\mathcal{F}}{\delta A_{\mu}}-\partial_{\mu}\xi\frac{\delta\mathcal{F}}{\delta B_{\mu}}+b\frac{\delta\mathcal{F}}{\delta \bar{c}}+\pi\frac{\delta\mathcal{F}}{\delta \bar{\xi}}+\frac{\delta \mathcal{F}}{\delta \bar{Y}}\frac{\delta\mathcal{F}}{\delta \psi}-\frac{\delta \mathcal{F}}{\delta Y}\frac{\delta\mathcal{F}}{\delta \bar{\psi}} \right]~,
\end{equation}
in the special case where $\mathcal{F}=\Gamma^{(0)}$, we recover the BRS invariance for the tree-level action:
\begin{equation}\label{STidentity}
\mathcal{S}(\Gamma^{(0)})|_{s=1}=0~.
\end{equation}
It is also useful to define the linearized Slavnov-Taylor operator ($\mathcal{S}_{\mathcal{F}}$): 
\begin{equation}\label{STL}
    \begin{aligned}
 \mathcal{S}_{\mathcal{F}}= &\int d^{4}x~\Big[ -\partial_{\mu}c\frac{\delta}{\delta A_{\mu}}-\partial_{\mu}\xi\frac{\delta}{\delta B_{\mu}}+b\frac{\delta}{\delta \bar{c}}+\pi\frac{\delta}{\delta \bar{\xi}}+\frac{\delta \mathcal{F}}{\delta \bar{Y}}\frac{\delta}{\delta \psi}+\frac{\delta \mathcal{F}}{\delta \psi}\frac{\delta}{\delta \bar{Y}} +\\
 &-\frac{\delta \mathcal{F}}{\delta Y}\frac{\delta}{\delta \bar{\psi}}-\frac{\delta \mathcal{F}}{\delta \bar{\psi}}\frac{\delta}{\delta Y} \Big]~,
\end{aligned}
\end{equation}
where from (\ref{ST}) and (\ref{STL}) it follows that the nilpotency identities are verified:
\begin{equation}\label{eqid1}
    \mathcal{S}_{\mathcal{F}}\mathcal{S}(\mathcal{F})=0~\forall\mathcal{F}~~~~\text{and}~~~~ \mathcal{S}_{\mathcal{F}}\mathcal{S}_{\mathcal{F}}=0~\text{if}~\mathcal{S}(\mathcal{F})=0~.
\end{equation}
Thanks to the Slavnov-Taylor identity $\mathcal{S}(\Gamma^{(0)})|_{s=1}=0$, the linearized Slavnov-Taylor operator is nilpotent, {\it i.e.} $(\mathcal{S}_{\Gamma^{(0)}})^2=0$. Nevertheless, the operator $\mathcal{S}_{\Gamma^{(0)}}$ acting on the fields and the antifields read: 
\begin{eqnarray}
&&\mathcal{S}_{\Gamma^{(0)}}\Phi=\mathfrak{s}\Phi~,~~\Phi=\{\psi,\bar{\psi},A_\mu,c,\bar{c},b,B_\mu,\xi,\bar{\xi},\pi\}~,\nonumber\\
&&\mathcal{S}_{\Gamma^{(0)}}Y=-\frac{\delta\Gamma^{(0)}}{\delta\bar{\psi}} ~,~~\mathcal{S}_{\Gamma^{(0)}}\bar{Y}=\frac{\delta\Gamma^{(0)}}{\delta\psi}~. \label{operation1}
\end{eqnarray}
Furthermore, the tree-level action $\Gamma^{(0)}$ satisfies the gauge conditions, ghost equations and the antighost equations as follows:
\begin{align}
&\frac{\delta \Gamma^{(0)}}{\delta b}=\partial^{\mu}A_{\mu}+\alpha b ~,~~ \frac{\delta \Gamma^{(0)}}{\delta \pi}=\partial^{\mu}B_{\mu}+\beta \pi~,\label{gaugecondi} \\
&\frac{\delta \Gamma^{(0)}}{\delta \bar{c}}=\Box c ~,~~ \frac{\delta \Gamma^{(0)}}{\delta \bar{\xi}}=\Box \xi~,\label{gh} \\
&\int d^{4}x\frac{\delta \Gamma^{(0)}}{\delta c}=\bar{\Delta}_{\text{I}} ~,~~ \int d^{4}x\frac{\delta \Gamma^{(0)}}{\delta \xi}=\bar{\Delta}_{\text{II}}~,\label{antgh}
\end{align}
where
\begin{align}
&\bar{\Delta}_{\text{I}}=\imath e\int d^{4}x[\bar{Y}\psi-\bar{\psi}Y]~,\\
&\bar{\Delta}_{\text{II}}=\imath g\int d^{4}x[\bar{Y}\psi-\bar{\psi}Y]~.
\end{align}

To conclude the symmetry analysis of dyon quantum electrodynamics (dQED), described by the classical action $\Gamma^{(0)}$ (\ref{eq35}), it shall be pointed out that the model is also invariant under two rigid symmetries which arising from the Slavnov-Taylor identity and the antighost equations, expressing the invariance under the gauge symmetry $U(1)\times U(1)$, indeed the conservation of electric charge and magnetic charge:
\begin{equation} 
\mathcal{W}_{\text{I}}^{\text{rig}}\Gamma^{(0)}=0 ~,~~ \mathcal{W}_{\text{II}}^{\text{rig}}\Gamma^{(0)}=0~,
\end{equation}
where the Ward operators associated to the two rigid symmetries are given by:
\begin{align}
&\mathcal{W}_{\text{I}}^{\text{rig}}=\imath e \int d^{4}x~\left[ \psi\frac{\delta}{\delta \psi}-\bar{\psi}\frac{\delta}{\delta \bar{\psi}}+Y\frac{\delta}{\delta Y}-\bar{Y}\frac{\delta}{\delta \bar{Y}} \right]~,  \label{rig1} \\
&\mathcal{W}_{\text{II}}^{\text{rig}}=\imath g \int d^{4}x~\left[ \psi\frac{\delta}{\delta \psi}-\bar{\psi}\frac{\delta}{\delta \bar{\psi}}+Y\frac{\delta}{\delta Y}-\bar{Y}\frac{\delta}{\delta \bar{Y}} \right]~.\label{rig2}
\end{align}

Additionally, from the Slavnov-Taylor operators, the gauge-fixing conditions, ghost and antighost equations, and rigid operators, we get the following operator closed algebra:
\begin{align}
&\frac{\delta}{\delta b}\mathcal{S}(\mathcal{F})-\mathcal{S}_{\mathcal{F}}\left(\frac{\delta\mathcal{F}}{\delta b}-\partial^{\mu}A_{\mu}\right)=\left(\frac{\delta \mathcal{F}}{\delta \bar{c}}-\Box c\right)~; \label{algebra1}\\
&\frac{\delta}{\delta \pi}\mathcal{S}(\mathcal{F})-\mathcal{S}_{\mathcal{F}}\left(\frac{\delta\mathcal{F}}{\delta \pi}-\partial^{\mu}B_{\mu}\right)=\left(\frac{\delta \mathcal{F}}{\delta \bar{\xi}}-\Box \xi\right)~;  \\
&\frac{\delta}{\delta \bar{c}}\mathcal{S}(\mathcal{F})+\mathcal{S}_{\mathcal{F}}\frac{\delta \mathcal{F}}{\delta \bar{c}}=0 ~;~~ \frac{\delta}{\delta \bar{\xi}}\mathcal{S}(\mathcal{F})+\mathcal{S}_{\mathcal{F}}\frac{\delta \mathcal{F}}{\delta \bar{\xi}}=0~; \\
&\int~d^{4}x~\frac{\delta}{\delta c}\mathcal{S}(\mathcal{F})+\mathcal{S}_{\mathcal{F}}\int~d^{4}x~\left(\frac{\delta \mathcal{F}}{\delta c}-\bar{\Delta}_{\text{I}}\right)=\mathcal{W}_{\text{I}}^{\text{rig}}(\mathcal{F})~;\\
&\int~d^{4}x~\frac{\delta}{\delta \xi}\mathcal{S}(\mathcal{F})+\mathcal{S}_{\mathcal{F}}\int~d^{4}x~\left(\frac{\delta \mathcal{F}}{\delta \xi}-\bar{\Delta}_{\text{II}}\right)=\mathcal{W}_{\text{II}}^{\text{rig}}(\mathcal{F})~;  \\
& \mathcal{W}_{\text{I}}^{\text{rig}}\mathcal{S}(\mathcal{F})-\mathcal{S}_{\mathcal{F}}\mathcal{W}_{\text{I}}^{\text{rig}}(\mathcal{F})=0 ~;~~ \mathcal{W}_{\text{II}}^{\text{rig}}\mathcal{S}(\mathcal{F})-\mathcal{S}_{\mathcal{F}}\mathcal{W}_{\text{II}}^{\text{rig}}(\mathcal{F})=0~,
\label{algebra6}
\end{align}
hence this closed operatorial algebra defines unequivocally the dyon quantum electrodynamics (dQED) model. Also, the equations above, (\ref{algebra1})--(\ref{algebra6}), can be used to analyze the model perturbatively, by looking for possible extensions of the model to the quantum level, through the use of the Becchi-Rouet-Stora (BRS) algebraic renormalization method together with the Quantum Action Principle~\cite{qap,piguet2}, with the Bogoliubov-Parasiuk-Hepp-Zimmermann-Lowenstein (BPHZL) renormalization procedure running in the background.

\section{Quantization}

\subsection{Searching for anomalies}

Now let us check the possibility of perturbatively extending the proposed model to the quantum level, thus to begin with, such an extension shall be described by a quantum vertex functional (or quantum action), denoted as $\Gamma\equiv\Gamma(A_{\mu},B_{\mu},\psi,\bar{\psi},c,\bar{c},b,\xi,\bar{\xi},\pi,Y,\bar{Y})$, satisfying the same constraints (\ref{algebra1})--(\ref{algebra6}) as the tree-level action $\Gamma^{(0)}$. The quantum action $\Gamma$ can be expanded as a formal power series in $\hbar$, in such a way that:
\begin{equation}\label{eqgama}
    \Gamma_{(s-1)}\equiv\Gamma=\Gamma^{(0)}+\mathcal{O}(\hbar)~,
\end{equation}
where the $\hbar$-zeroth order action, $\Gamma^{(0)}$, coincides with the classical action.

A classical symmetry might not survive at the quantum level, so in the case which a classical symmetry is broken at the quantum level, we call this breaking an anomaly. Some anomalies could not compromise necessarily the consistency of the model, once they simply express that the corresponding symmetries do not hold anymore at the quantum level, however there are those ones that spoil the spectrum consistency of the model. Generally, in the study of anomalies, we may encounter what is referred to as a consistent anomaly, and as any anomaly, it is independent of any regularization or renormalization scheme. Such an anomaly acts as an obstruction to the model at the quantum level, by compromising its spectrum consistency. This, in turn, leads to a violation of $S$-matrix unitarity, as demonstrated by Kugo and Ojima~\cite{kugo}. 

A consistent anomaly might arise from the Slavnov-Taylor identity if it is violated at the quantum level. Supposing that, at the order $\hbar^{(n-1)}$, the Slavnov-Taylor identity is preserved, then the breakdown shows appears at the order $\hbar^n$, therefore, according to the Quantum Action Principle~\cite{qap,piguet2}, it follows that:
\begin{equation}\label{eqqpa}
    \mathcal{S}(\Gamma)\Big|_{s=1}=\hbar^{n} \Delta\cdot\Gamma\Big|_{s=1}=\hbar^{n}\Delta+\mathcal{O}(\hbar^{n+1})~,
\end{equation}
here, $\Delta\equiv\Delta|_{s=1}$ is an integrated local polynomial in the fields and external sources, with ghost number $1$,  and bounded by UV dimension, $d\leq 4$, and IR dimension, $r\geq 4$ (see Table \ref{tab1}).

Now, by combining the expansion of the quantum action \eqref{eqgama} with the linearized Slavnov-Taylor operator \eqref{STL}:
\begin{equation}
    \mathcal{S}_{\Gamma}=\mathcal{S}_{\Gamma^{(0)}}+\mathcal{O}(\hbar)~,
\end{equation}
then it follows, from the nilpotency \eqref{eqid1} and the Slavnov-Taylor quantum breaking \eqref{eqqpa}, a consistency condition for the quantum breaking ($\Delta$):
\begin{equation} \label{wzcondition}
    \mathcal{S}_{\Gamma^{(0)}}\Delta=0~,
\end{equation}
the well-known Wess-Zumino consistency condition.

Additionally, other than the Wess-Zumino consistency condition \eqref{wzcondition}, by taking into consideration the Slavnov-Taylor identity \eqref{STidentity}, the gauge \eqref{gaugecondi}, the ghost \eqref{gh} and antighost \eqref{antgh} equations, and the two rigid conditions \eqref{rig1}, the Slavnov-Taylor quantum breaking \eqref{eqqpa} must fulfill the following constraints:
\begin{eqnarray}
&&\frac{\delta\Delta}{\delta b}=\frac{\delta\Delta}{\delta \bar{c}}=\int d^{4}x~\frac{\delta \Delta}{\delta c}= \mathcal{W}_{\text{I}}^{\text{rig}}\Delta=0~;\label{constrdelta1} \\ 
&&\frac{\delta\Delta}{\delta \pi}=\frac{\delta\Delta}{\delta \bar{\xi}}=\int d^{4}x~\frac{\delta \Delta}{\delta \xi}=\mathcal{W}_{\text{II}}^{\text{rig}}\Delta=0~. \label{constrdelta2}
\end{eqnarray}

These constraints lead us to the conclusion that the local polynomial, $\Delta$, whose ghost number is one, depends explicitly only in terms of $A_{\mu}$, $B_{\mu}$, $\partial^{\mu}c$, and $\partial^{\mu}\xi$. It is important to point out that, regarding the invariance of $\Delta$ under the rigid symmetries $\mathcal{W}_{\text{I}}^{\text{rig}}$ and $\mathcal{W}_{\text{II}}^{\text{rig}}$, as outlined in \eqref{constrdelta1} and \eqref{constrdelta2}, some subtleties about these invariances should be made explicit. This is because the symmetry group $U(1)\times U(1)$ is a non-semisimple Lie group, which implies that the associated rigid symmetry might, in principle, be anomalous. However, as long as both Abelian factors are not spontaneously broken and the electric charge ($e$) and magnetic charge ($g$) are conserved, the conditions $\mathcal{W}_{\text{I}}^{\text{rig}}\Gamma^{(0)}=0$ and $\mathcal{W}_{\text{II}}^{\text{rig}}\Gamma^{(0)}=0$ hold true. Consequently, the conditions $\mathcal{W}_{\text{I}}^{\text{rig}}\Delta=0$ and $\mathcal{W}_{\text{II}}^{\text{rig}}\Delta=0$ are still fulfilled~\cite{stora,kraus}.

The Wess-Zumino consistency condition corresponds to a cohomology problem in the sector of ghost number one, bounded by $d \leq 4$ and $r \geq 4$. The general solution is written as:
\begin{equation}
\Delta = \widetilde{\Delta} + \mathcal{S}_{\Gamma^{(0)}}\widehat{\Delta}~,
\end{equation}
where $\widetilde{\Delta}$ is a non-trivial term with ghost number $1$, while $\mathcal{S}_{\Gamma^{(0)}}\widehat{\Delta}$ is a trivial cocycle, with $\widehat{\Delta}$ having ghost number $0$. The non-trivial cocycle, $\widetilde{\Delta}$, represents a true anomaly in the Slavnov-Taylor identity, whereas $\widehat{\Delta}$ corresponds to the so-called non-invariant counterterms, which should be absorbed into the quantum action order by order. From the constraints \eqref{constrdelta1} and \eqref{constrdelta2}:
\begin{equation}
    \int d^{4}x~\frac{\delta \Delta}{\delta c}=0 ~~~~\text{and}~~~~ \int d^{4}x~\frac{\delta \Delta}{\delta \xi}=0~,
\end{equation}
therefore, it follows that $\Delta$ can be expressed as:
\begin{equation}\label{eqdelta}
    \Delta=\int d^{4}x~[\mathcal{K}_{\mu}\partial^{\mu}c+\mathcal{R}_{\mu}\partial^{\mu}\xi]~,
\end{equation}
where $\mathcal{K}_{\mu}$ and $\mathcal{R}_{\mu}$ are rank one tensor bounded by UV and IR dimension, $d\leq 3$ and $r\geq 3$, respectively, with ghost number $0$. In addition to, they can be rewritten as follows:
\begin{equation}\label{KV}
\displaystyle\mathcal{K}_{\mu}=\sum_{k}\alpha_{k}\mathcal{V}_{\mu}^{k} ~~~~\text{and}~~~~ \displaystyle\mathcal{R}_{\mu}=\sum_{k}\beta_{k}\mathcal{V}_{\mu}^{k}~,
\end{equation}
with $\alpha_{k}$ and $\beta_{k}$ being arbitrary scalar coefficients. The list below contains all the candidates for vectors $\mathcal{V}_{\mu}^{k}$: 
\begin{equation}
    \begin{matrix}
    \mathcal{V}_{\mu}^{1}=A_{\mu}A_{\nu}A^{\nu}~;&\mathcal{V}_{\mu}^{2}=A_{\mu}A_{\nu}B^{\nu}~;  &\mathcal{V}_{\mu}^{3}=A_{\mu}B_{\nu}B^{\nu}~; \\ 
    \mathcal{V}_{\mu}^{4}=B_{\mu}B_{\nu}B^{\nu}~;&\mathcal{V}_{\mu}^{5}=B_{\mu}B_{\nu}A^{\nu}~;  &\mathcal{V}_{\mu}^{6}=B_{\mu}A_{\nu}A^{\nu}~; \\ 
    \mathcal{V}_{\mu}^{7}=\partial_{\mu}A_{\nu}A^{\nu}~;&\mathcal{V}_{\mu}^{8}=\partial_{\mu}A_{\nu}B^{\nu}~;  &\mathcal{V}_{\mu}^{9}=\partial_{\mu}B_{\nu}A^{\nu}~; \\ 
    \mathcal{V}_{\mu}^{10}=\partial_{\mu}B_{\nu}B^{\nu}~;&\mathcal{V}_{\mu}^{11}=\partial_{\nu}A^{\nu}A_{\mu}~;  &\mathcal{V}_{\mu}^{12}=\partial_{\nu}A^{\nu}B_{\mu}~; \\ 
    \mathcal{V}_{\mu}^{13}=\partial_{\nu}B^{\nu}A_{\mu}~;&\mathcal{V}_{\mu}^{14}=\partial_{\nu}B^{\nu}B_{\mu}~;  &\mathcal{V}_{\mu}^{15}=A_{\nu}\partial^{\nu}A_{\mu}~; \\ 
    \mathcal{V}_{\mu}^{16}=A_{\nu}\partial^{\nu}B_{\mu}~;&\mathcal{V}_{\mu}^{17}=B_{\nu}\partial^{\nu}A_{\mu}~;  &\mathcal{V}_{\mu}^{18}=B_{\nu}\partial^{\nu}B_{\mu}~; \\ 
    \mathcal{V}_{\mu}^{19}=\varepsilon_{\mu\nu\rho\sigma}\partial^{\nu}A^{\rho}B^{\sigma}~;&\mathcal{V}_{\mu}^{20}=\varepsilon_{\mu\nu\rho\sigma}\partial^{\nu}A^{\rho}A^{\sigma}~;  &\mathcal{V}_{\mu}^{21}=\varepsilon_{\mu\nu\rho\sigma}\partial^{\nu}B^{\rho}B^{\sigma}~;  \\
    \mathcal{V}_{\mu}^{22}=\partial_{\mu}\partial^{\nu}B_{\nu}~;&\mathcal{V}_{\mu}^{23}=\partial_{\mu}\partial^{\nu}A_{\nu}~.
\end{matrix}
\end{equation}

It is worth noting that the Slavnov-Taylor operator, $\mathcal{S}(\mathcal{F})$, as well as the linearized Slavnov-Taylor operator, $\mathcal{S}_{\mathcal{F}}$, are invariant under the discrete transformations ${\widetilde C}$ \eqref{charge}, $P$ \eqref{parity}, and $T$ \eqref{time}, provided that the functional $\mathcal{F}$ itself be invariant under these discrete transformations.

Due to the invariance of the action $\Gamma^{(0)}$ together with the fact that the Slavnov-Taylor operator is even under the discrete transformations ${\widetilde C}$, $P$, and $T$, it follows that the possible anomaly term $\Delta$ must also be even under these symmetries. Also, from \eqref{eqdelta}, since $\partial^{\mu}c$ and $\partial^{\mu}\xi$ are odd under charge conjugation (${\widetilde C}$), it implies that $\mathcal{K}{\mu}$ and $\mathcal{R}{\mu}$ must be odd under ${\widetilde C}$. Consequently, we eliminate from the basis of vectors ${\mathcal{V}^{k}_{\mu}}$ the elements corresponding to $k=7,\ldots,21$, then the remaining read:
\begin{equation}\label{lista2}
    \begin{matrix}
    \mathcal{V}_{\mu}^{1}=A_{\mu}A_{\nu}A^{\nu}~;&\mathcal{V}_{\mu}^{2}=A_{\mu}A_{\nu}B^{\nu}~;  &\mathcal{V}_{\mu}^{3}=A_{\mu}B_{\nu}B^{\nu}~; \\ 
    \mathcal{V}_{\mu}^{4}=B_{\mu}B_{\nu}B^{\nu}~;&\mathcal{V}_{\mu}^{5}=B_{\mu}B_{\nu}A^{\nu}~;  &\mathcal{V}_{\mu}^{6}=B_{\mu}A_{\nu}A^{\nu}~;    \\
    \mathcal{V}_{\mu}^{22}=\partial_{\mu}\partial^{\nu}B_{\nu}~;&\mathcal{V}_{\mu}^{23}=\partial_{\mu}\partial^{\nu}A_{\nu}~.
    \end{matrix}
\end{equation}
With respect to parity ($P$), $\partial^{\mu}c$ is a vector, while $\partial^{\mu}\xi$ is a pseudovector, as a consequence $\mathcal{K}{\mu}$ is a vector, and $\mathcal{R}{\mu}$ is a pseudovector, so it follows that:
\begin{equation}
\mathcal{K}_{\mu}=\{\mathcal{V}^{1}_{\mu}, \mathcal{V}^{3}_{\mu}, \mathcal{V}^{5}_{\mu},\mathcal{V}_{\mu}^{23}\} ~~~~\text{and}~~~~ \mathcal{R}_{\mu}=\{\mathcal{V}^{2}_{\mu}, \mathcal{V}^{4}_{\mu}, \mathcal{V}^{6}_{\mu},\mathcal{V}_{\mu}^{22}\}~,
\end{equation}
and by recalling \eqref{eqdelta} and \eqref{KV}, the possible anomaly $\Delta$ that fulfills all discrete symmetries, namely ${\widetilde C}$, $P$, and $T$, can be written as:
\begin{equation}\label{eqdelta1}
	\begin{split}
		\Delta &=\int d^{4}x\Big[ \alpha_{1}A_{\mu}A_{\nu}A^{\nu}\partial^{\mu}c+
                \alpha_{3}A_{\mu}B_{\nu}B^{\nu}\partial^{\mu}c+ 
                \alpha_{5}B_{\mu}B_{\nu}A^{\nu}\partial^{\mu}c+\alpha_{23}\partial_{\mu}\partial^{\nu}A_{\nu}\partial^{\mu}c +\\
		&+\beta_{2}A_{\mu}A_{\nu}B^{\nu}\partial^{\mu}\xi+ 
                \beta_{4}B_{\mu}B_{\nu}B^{\nu}\partial^{\mu}\xi+
                \beta_{6}B_{\mu}A_{\nu}A^{\nu}\partial^{\mu}\xi+\beta_{22}\partial_{\mu}\partial^{\nu}B_{\nu}\partial^{\mu}\xi
                \Big]~.
	\end{split}    
\end{equation}

Furthermore, bearing in mind the action of the linearized Slavnov-Taylor, $\mathcal{S}_{\Gamma^{(0)}}$ \eqref{operation1}, we verify that the expression above for the possible anomaly, $\Delta$, can be rewritten as:
\begin{equation}
\Delta=\mathcal{S}_{\Gamma^{(0)}}(\lambda_{1}\widehat{\Delta}_{1}+\lambda_{2}\widehat{\Delta}_{2}+\lambda_{3}\widehat{\Delta}_{3}+\lambda_{4}\widehat{\Delta}_{4}+\lambda_{5}\widehat{\Delta}_{5}+\lambda_{6}\widehat{\Delta}_{6})~,
\end{equation}
so that the integrated local monomials $\widehat{\Delta}_{k}$ $(k=1,\ldots,6)$, are given by:
\begin{align}
&\widehat{\Delta}_{1}=\int d^{4}x(A_{\mu}A^{\mu})^{2}~;~~\widehat{\Delta}_{2}=\int d^{4}x(B_{\mu}B^{\mu})^{2}~;\\
&\widehat{\Delta}_{3}=\int d^{4}x(A_{\mu}A^{\mu}B_{\nu}B^{\nu})~;~~\widehat{\Delta}_{4}=\int d^{4}x(A_{\mu}A^{\nu}B_{\nu}B^{\mu})~;\\
&\widehat{\Delta}_{5}=\int d^{4}x(\partial_{\mu}\partial^{\nu}B_{\nu}B^{\mu})~;~~\widehat{\Delta}_{6}=\int d^{4}x(\partial_{\mu}\partial^{\nu}A_{\nu}A^{\mu})~,
\end{align}
with the coefficients $\lambda_{i}$ being related to the coefficients $\alpha_{k}$ and $\beta_{k}$ as
\begin{equation}
\lambda_{1}=-\frac{\alpha_{1}}{4}~,~~\lambda_{2}=-\frac{\beta_{4}}{4}~,~~\lambda_{3}=-\frac{\alpha_{3}}{2}=-\frac{\beta_{6}}{2}
~,~~\lambda_{4}=-\frac{\alpha_{5}}{2}=-\frac{\beta_{2}}{2}~,~~\lambda_{5}=-\frac{\alpha_{23}}{2}~,~~\lambda_{6}=-\frac{\beta_{22}}{2}~.
\end{equation}

Finally, we demonstrate that, with respect to the anomaly $\Delta = \widetilde{\Delta} + \mathcal{S}_{\Gamma^{(0)}}\widehat{\Delta}$, we get $\widetilde{\Delta} = 0$, so there is no gauge anomaly, as a consequence, the only contribution comes from the non-invariant counterterms, $\widehat{\Delta}$, which shall be incorporated into the quantum action order by order: 
\begin{equation} 
\mathcal{S}_{\Gamma^{(0)}}(\Gamma-\hbar^n{\widehat{\Delta}}) \equiv \mathcal{S}_{\Gamma^{(0)}}\left(\Gamma-\hbar^n\lambda_{1}\widehat{\Delta}_{1}-\hbar^n\lambda_{2}\widehat{\Delta}_{2}-\hbar^n\lambda_{3}\widehat{\Delta}_{3}-\hbar^n\lambda_{4}\widehat{\Delta}_{4}-\hbar^n\lambda_{5}\widehat{\Delta}_{5}-\hbar^n\lambda_{6}\widehat{\Delta}_{6}\right) = 0\hbar^n + {\mathcal{O}}(\hbar^{n+1})~. \label{a_final}
\end{equation}
In addition to, it should be stressed that, although there is no gauge anomaly, the model is also free from infrared anomalies, which would be originated -- thanks to the presence massless gauge fields -- if at least one of the non-invariant counterterms, $\widehat{\Delta}_{k}$, had violated the infrared condition $r \geq 4$.

\subsection{Searching for counterterms}

The multiplicative renormalizability, also known as the stability condition, is accomplished whenever perturbative quantum corrections produce just local counterterms corresponding to renormalization of parameters which are already present in the tree-level action, accordingly those radiative corrections shall be reabsorbed order by order through redefinitions of the classical quantities, {\it e.g.} fields, coupling constants and masses. Consequently, in order to check if the tree-level action $\Gamma^{(0)}$ \eqref{eq35} is multiplicative renormalizable, it is perturbed by an arbitrary integrated local functional in the fields, {\it i.e.} the counterterm $\Gamma^{\rm c} \equiv \Gamma^{\rm c}_{(s-1)}$: 
\begin{equation}
    \Gamma^{(0)}\to\Gamma^{(0)}+\varepsilon\Gamma^{\rm c}~,
\end{equation}
where $\varepsilon$ is an infinitesimal parameter, and the counterterm action ($\Gamma^{\rm c}$) has the same quantum numbers as the tree-level action ($\Gamma^{(0)}$). Once the perturbed action must satisfy the same conditions as the initial action, taking into consideration that $\mathfrak{s}\Gamma^{(0)}=(s-1)\Delta_{\rm lin}$, we get:
\begin{align}
&\mathcal{S}(\Gamma^{(0)}+\varepsilon\Gamma^{\rm c})=(s-1)\Delta_{\rm lin} ~~\therefore~~ \mathcal{S}(\Gamma^{(0)})+\varepsilon\mathcal{S}_{\Gamma^{(0)}}(\Gamma^{\rm c})=(s-1)\Delta_{\rm lin} ~~\therefore \nonumber\\
&(s-1)\Delta_{\rm lin}+\varepsilon\mathcal{S}_{\Gamma^{(0)}}(\Gamma^{\rm c})=(s-1)\Delta_{\rm lin} ~~\therefore~~ \mathcal{S}_{\Gamma^{(0)}}\Gamma^{\rm c}=0~,
\end{align}
hence $\Gamma^{\rm c}$ remains BRS invariant, even though the initial action ($\Gamma^{(0)}$) exhibits a linear breaking ($\Delta_{\rm lin}$) due to the Lowenstein-Zimmermann mass term action, $\Sigma_{\rm IR}$ \eqref{IR}.

Analogously to the line of action presented above, since the perturbed action must satisfy the same conditions as the initial action, thus from the identities \eqref{gaugecondi}--\eqref{rig2}, we get the following constraints for $\Gamma^{\rm c}$:
\begin{align}
&\displaystyle\mathcal{S}_{\Gamma^{(0)}}\Gamma^{\rm c}=0~,\label{stbwz}\\
&\displaystyle\frac{\delta \Gamma^{\rm c}}{\delta b}=\displaystyle\frac{\delta \Gamma^{\rm c}}{\delta \bar{c}}=\displaystyle\frac{\delta \Gamma^{\rm c}}{\delta c}=\displaystyle\mathcal{W}_{\rm I}^{\rm rig}\Gamma^{\rm c}=0~,\label{constr1}\\
&\displaystyle\frac{\delta \Gamma^{\rm c}}{\delta \pi}=\displaystyle\frac{\delta \Gamma^{\rm c}}{\delta \bar{\xi}}=\displaystyle\frac{\delta \Gamma^{\rm c}}{\delta \xi}=\displaystyle\mathcal{W}_{\rm II}^{\rm rig}\Gamma^{\rm c}=0~,\label{constr2}
\end{align}
beyond that, the counterterm action $\Gamma^{\rm c}$ has to be invariant under the discrete transformations ${\widetilde C}$ \eqref{charge}, $P$ \eqref{parity}, and $T$ \eqref{time}:
\begin{equation}\label{constr3}
\Gamma^{\rm c}\overset{{\widetilde C}}{\longrightarrow}\Gamma^{\rm c}~,~~\Gamma^{\rm c}\overset{P}{\longrightarrow}\Gamma^{\rm c}~,~~\Gamma^{\rm c}\overset{T}{\longrightarrow}\Gamma^{\rm c}~,~~\Gamma^{\rm c}\overset{{\widetilde C}PT}{\longrightarrow}\Gamma^{\rm c}~.
\end{equation}

It is appropriate to point out that, the identity $\mathcal{S}_{\Gamma^{(0)}}\Gamma^{\rm c}=0$ \eqref{stbwz} reminds the same problem analyzed in the previous section, $\mathcal{S}_{\Gamma^{(0)}}\Delta=0$ \eqref{wzcondition}, a cohomology problem, albeit being in sector ghost number zero, with UV and IR dimensions bounded by $d \leq 4$ and $r \geq 4$, respectively: 
\begin{equation}\label{equacaodeestabilidade}             
\Gamma^{\rm c}=\widetilde{\Gamma}+\mathcal{S}_{\Gamma^{(0)}}\widehat{\Gamma}~.
\end{equation}
In this case, $\widetilde{\Gamma}$ represents the non-trivial term with ghost number $0$, while $\widehat{\Gamma}$ is the trivial cocycle with ghost number $-1$. The counterterms from the non-trivial part can be reabsorbed through the renormalization of the gauge coupling constants and mass, which are the physical parameters. On the other hand, the trivial cocycle counterterms can be reabsorbed through a redefinition of the field amplitudes and gauge parameters, indicating a non-physical renormalization corresponding to the cohomological triviality of $\mathcal{S}_{\Gamma^{(0)}}\widehat{\Gamma}$.

The counterterm ($\Gamma^{\rm c}$), which has ghost number zero and fulfill the UV and IR constraints, $d \leq 4$ and $r \geq 4$, respectively, can be written in a basis of local monomials as follows:
\begin{equation} \label{mono}
\Gamma^{\rm c}=\int~d^{4}x~\sum_{i}\alpha_{i}\Sigma^{i}~,
\end{equation}
where the local monomials $\Sigma^{i}$, according to the constraints \eqref{constr1}--\eqref{constr2}, are listed below:
\begin{equation}
    \begin{matrix}
        \Sigma^{1}=\imath\bar{\psi}\gamma^{\mu}\partial_{\mu}\psi~,~~&\Sigma^{2}=\bar{\psi}\gamma^{\mu}\psi A_{\mu}~,~~ &\Sigma^{3}=\bar{\psi}\gamma^{\mu}\psi B_{\mu}~, \\ 
        \Sigma^{4}=m\bar{\psi}\psi~,~~&\Sigma^{5}=A_{\mu}A^{\mu}A_{\nu}A^{\nu}~,~~&\Sigma^{6}=A_{\mu}A^{\mu}A_{\nu}B^{\nu}~, \\
        \Sigma^{7}=A_{\mu}A^{\mu}B_{\nu}B^{\nu}~,~~&\Sigma^{8}=B_{\mu}B^{\mu}B_{\nu}B^{\nu}~,~~&\Sigma^{9}=B_{\mu}B^{\mu}A_{\nu}B^{\nu}~, \\ 
        \Sigma^{10}=A_{\mu}B^{\mu}A_{\nu}B^{\nu}~,~~&\Sigma^{11}=\partial_{\mu}A_{\nu}A^{\mu}B^{\nu}~,~~&\Sigma^{12}=\partial_{\mu}A_{\nu}B^{\mu}B^{\nu}~, \\ 
        \Sigma^{13}=\partial_{\mu}A^{\mu}B_{\nu}B^{\nu}~,~~&\Sigma^{14}=\partial_{\mu}B^{\mu}B_{\nu}B^{\nu}~,~~&\Sigma^{15}=\partial_{\mu}B_{\nu}A^{\mu}A^{\nu}~, \\ 
        \Sigma^{16}=\partial_{\mu}B_{\nu}A^{\nu}B^{\mu}~,~~&\Sigma^{17}=\partial_{\mu}B_{\nu}A^{\mu}B^{\nu}~,~~&\Sigma^{18}=\partial_{\mu}B_{\nu}B^{\mu}B^{\nu}~, \\ 
        \Sigma^{19}=\partial_{\mu}A^{\mu}A_{\nu}A^{\nu}~,~~&\Sigma^{20}=\partial_{\mu}B^{\mu}A_{\nu}A^{\nu}~,~~&\Sigma^{21}=\partial_{\mu}A_{\nu}A^{\mu}A^{\nu}~, \\ 
        \Sigma^{22}=\partial_{\mu}A_{\nu}B^{\mu}A^{\nu}~,~~&\Sigma^{23}=\partial_{\mu}A_{\nu}\partial^{\mu}A^{\nu}~,~~&\Sigma^{24}=\partial_{\mu}A_{\nu}\partial^{\mu}B^{\nu}~, \\ 
        \Sigma^{25}=\partial_{\mu}B_{\nu}\partial^{\mu}B^{\nu}~,~~&\Sigma^{26}=\partial_{\mu}A_{\nu}\partial^{\nu}A^{\mu}~,~~&\Sigma^{27}=\partial_{\mu}A_{\nu}\partial^{\nu}B^{\mu}~, \\
        \Sigma^{28}=\partial_{\mu}B_{\nu}\partial^{\nu}B^{\mu}~,~~&\Sigma^{29}=\varepsilon^{\mu\nu\rho\sigma}A_{\mu}A_{\nu}A_{\rho}A_{\sigma}~,~~&\Sigma^{30}=\varepsilon^{\mu\nu\rho\sigma}A_{\mu}A_{\nu}A_{\rho}B_{\sigma}~, \\
        \Sigma^{31}=\varepsilon^{\mu\nu\rho\sigma}A_{\mu}A_{\nu}B_{\rho}B_{\sigma}~,~~&\Sigma^{32}=\varepsilon^{\mu\nu\rho\sigma}A_{\mu}B_{\nu}B_{\rho}B_{\sigma}~,~~&\Sigma^{33}=\varepsilon^{\mu\nu\rho\sigma}B_{\mu}B_{\nu}B_{\rho}B_{\sigma}~,
                 \\
        \Sigma^{34}=\varepsilon^{\mu\nu\rho\sigma}\partial_{\mu}A_{\nu}A_{\rho}A_{\sigma}~,~~&\Sigma^{35}=\varepsilon^{\mu\nu\rho\sigma}\partial_{\mu}A_{\nu}A_{\rho}B_{\sigma}~,~~&\Sigma^{36}=\varepsilon^{\mu\nu\rho\sigma}\partial_{\mu}A_{\nu}B_{\rho}B_{\sigma}~,
                 \\
        \Sigma^{37}=\varepsilon^{\mu\nu\rho\sigma}\partial_{\mu}B_{\nu}B_{\rho}B_{\sigma}~,~~&\Sigma^{38}=\varepsilon^{\mu\nu\rho\sigma}\partial_{\mu}A_{\nu}\partial_{\rho}B_{\sigma}~,~~&\Sigma^{39}=\Box\partial_{\mu}A^{\mu}~,
                \\
        \Sigma^{40}=\Box\partial_{\mu}B^{\mu}~.&&
                 
    \end{matrix}
\end{equation}

Since the counterterm, $\Gamma^{\rm c}$ \eqref{mono}, shall also to be invariant under charge conjugation (${\widetilde C}$), parity ($P$) and time reversal ($T$) transformations, the monomials $\Sigma^{i}$ associated to $i=(6,9,11,\ldots,22,24,27,29,\ldots,40)$ vanish necessarily, thus it remains: 
\begin{equation}\label{eqgamma3}
	\begin{split}
		\Gamma^{\rm c} &=\int d^{4}x\Big[ \alpha_{1}\imath\bar{\psi}\gamma^{\mu}\partial_{\mu}\psi+
                \alpha_{2}\bar{\psi}\gamma^{\mu}\psi A_{\mu}+ 
                \alpha_{3}\bar{\psi}\gamma^{\mu}\psi B_{\mu}+
                \alpha_{4}m\bar{\psi}\psi+
                \\
		&+
		\alpha_{5}A_{\mu}A^{\mu}A_{\nu}A^{\nu}+
                \alpha_{7}A_{\mu}A^{\mu}B_{\nu}B^{\nu}+
                \alpha_{8}B_{\mu}B^{\mu}B_{\nu}B^{\nu}+ 
                \alpha_{10}A_{\mu}B^{\mu}A_{\nu}B^{\nu}+\\
            &+ 
                \alpha_{23}\partial_{\mu}A_{\nu}\partial^{\mu}A^{\nu}+
                \alpha_{25}\partial_{\mu}B_{\nu}\partial^{\mu}B^{\nu}+
                \alpha_{26}\partial_{\mu}A_{\nu}\partial^{\nu}A^{\mu}+
                \alpha_{28}\partial_{\mu}B_{\nu}\partial^{\nu}B^{\mu}
                \Big]~.
	\end{split}    
\end{equation}
Now, bearing in mind that the counterterm has also to satisfy the Slavnov-Taylor invariance condition, $\mathcal{S}{\Gamma^{(0)}}\Gamma^{\rm c} = 0$ \eqref{stbwz}, it results that the coefficients $\alpha_5=\alpha_7=\alpha_8=\alpha_{10}=0$, while the remaining coefficients are redefined as:  
\begin{equation}
\alpha_{2}=e\alpha_{1}~,~~\alpha_{3}=g\alpha_{1}~,~~\alpha_{23}=-\alpha_{26}=-\frac{\beta_{1}}{2}~,~~\alpha_{25}=-\alpha_{28}=-\frac{\beta_{2}}{2}~,
\end{equation}
as a consequence, in accordance with the constraints \eqref{stbwz}--\eqref{constr3}, along with UV and IR dimensions bounded by $d \leq 4$ and $r \geq 4$, respectively, the action of the invariant counterterm ($\Gamma^{\rm c}$) reads:
\begin{equation}\label{eqgamma4}
\Gamma^{\rm c}=\int d^{4}x\left[ \alpha_{1}\imath\bar{\psi}\gamma^{\mu}D_{\mu}\psi+\alpha_{4}m\bar{\psi}\psi-\frac{1}{4}\beta_{1}F_{\mu\nu}F^{\mu\nu}-\frac{1}{4}\beta_{2}G_{\mu\nu}G^{\mu\nu} \right]~.
\end{equation}
Nevertheless, thanks to the duality symmetry of the fields and parameters, presented in \eqref{trocatrocafields}, we get $\beta_{1}=\beta_{2}=\beta$, thereby:
\begin{equation}\label{final_counterterm}
\Gamma^{\rm c}=\int d^{4}x\left\{ \alpha_{1}\imath\bar{\psi}\gamma^{\mu}D_{\mu}\psi+\alpha_{4}m\bar{\psi}\psi-\frac{1}{4}\beta \left[ F_{\mu\nu}F^{\mu\nu} + G_{\mu\nu}G^{\mu\nu}\right] \right\}~,
\end{equation}
where the coefficients, $\alpha_{1}$, $\alpha_{4}$, and $\beta$, are fixed by the following renormalization conditions:
\begin{align}
&\displaystyle\frac{d}{d\slashed{p}}\Gamma_{\bar{\psi}\psi}(\slashed{p})\Big|_{\slashed{p}=m}=1~,~~\displaystyle\Gamma_{\bar{\psi}\psi}(\slashed{p})\Big|_{\slashed{p}=m}=0~,  \\
&\displaystyle\frac{\partial}{\partial p^{2}}\Gamma^{\rm T}_{AA}(p^{2})\Big|_{p^{2}=\mu^{2}}=\displaystyle\frac{\partial}{\partial p^{2}}\Gamma^{\rm T}_{BB}(p^{2})\Big|_{p^{2}=\mu^{2}}=1~,
\end{align}
with $\mu$ being an energy scale. Ultimately, we conclude that the model is multiplicatively renormalizable at all orders in perturbation theory.

In summary, by taking into consideration the former result \eqref{a_final}, on searching for anomalies, together with the last one \eqref{final_counterterm}, on searching for counterterms, completes the proof about the quantum consistency at all orders in perturbation theory of the dyon quantum electrodynamics (dQED).

\section{Conclusions}

The dyon quantum electrodynamics (dQED), a Cabibbo-Ferrari~\cite{cabibbo1962quantum} and Salam~\cite{salam1966magnetic} based model, which describes an alternative approach to implementing fundamental magnetic monopoles, by breaking the Dirac quantization paradigm, has been analyzed in details. The BRS symmetry of the model was established, along with the gauge-fixing and its discrete symmetries, namely charge conjugation, parity and time reversal, and the duality symmetry of the fields and parameters. The tree-level propagators were computed, the spectrum consistency (causality and unitarity) has been verified, concluding that the dQED model is free from tachyons and ghosts, besides it was also conjectured that the Froissart-Martin unitarity bound~\cite{itzykson,chaichian} shall not be jeopardized. The BRS invariance was expressed in a functional way through the Slavnov-Taylor identity, together with the gauge-fixing conditions, ghost and antighost equations, and the two rigid identities. Ultimately, we prove that the dQED model is multiplicative renormalizable and free from any kind of anomaly (gauge, rigid and infrared), that being so, the model is fully renormalizable at all orders in perturbation theory. It should be mentioned that, since the BRS algebraic renormalization method~\cite{becchi,piguet-rouet} and the BPHZL subtraction scheme~\cite{Low} are derived from general theorems of perturbative quantum field theory~\cite{qap,piguet-rouet}, the proof presented here is independent of any regularization scheme.

\subsection*{Acknowledgements}
The CAPES-Brazil is acknowledged for invaluable financial help.



\begin{thebibliography}{00}
\bibitem{dirac1931} P.A.M. Dirac, Proc. Roy. Soc. Lond. A133 (1931) 60.

\bibitem{cabibbo1962quantum} N. Cabibbo, E. Ferrari, Il Nuovo Cimento 23 (1962) 1147

\bibitem{salam1966magnetic} A. Salam, Phys. Lett. 22 (1966) 683.

\bibitem{Schwinger1968} J. Schwinger, Phys. Rev. 173 (1968) 1536.

\bibitem{thooftPolyakov} G. 't Hooft, Nucl. Phys. B79 (1974) 276; A.M. Polyakov, JETP Lett. 20 (1974) 194.

\bibitem{nambu197476} Y. Nambu, Phys. Rev. D12 (1974) 4262; Y. Nambu, Nucl. Phys. B130 (1977) 505.

\bibitem{Azevedodaniel} D.O.R. Azevedo, M.L. Bispo, O.M. Del Cima, D.H.T. Franco, A.R. Pereira, EPL 136 (2021) 30004.

\bibitem{dirac1948} P.A.M. Dirac, Phys. Rev. 74 (1948) 817.

\bibitem{deliyergiyev2016recent} M. Deliyergiyev, Open Phys. 14 (2016) 281.

\bibitem{bento2024classes} M.P. Bento, H.E. Haber, J.P. Silva, Phys. Lett. B850 (2024) 138501.

\bibitem{Bauer:2018onh} M. Bauer, P. Foldenauer, J. Jaeckel, J. High Energy Phys. (JHEP) 07 (2018) 094.

\bibitem{ramsey1958time} N.F. Ramsey, Phys. Rev. 109 (1958) 225.

\bibitem{rajantie2016magnetic} A. Rajantie, Phys. Today 69 (2016) 40.

\bibitem{Rabl:1969gx} A. Rabl, Phys. Rev. 179 (1969) 1363.

\bibitem{Datta:1983um} T. Datta, Lett. Nuovo Cim. 37 (1983) 51.

\bibitem{singleton1996does} D. Singleton, Int. J. Theor. Phys. 35 (1996) 2419.

\bibitem{Govaerts:2023iqf} J. Govaerts, Eur. Phys. J. C83 (2023) 158.

\bibitem{msc-thadeu} T.S. Dias, {\it The quantum Cabibbo-Ferrari-Salam electrodynamics}, M.Sc. thesis, advisor: O.M. Del Cima, Universidade Federal de Vi\c cosa (UFV), Vi\c cosa, Minas Gerais, Brazil, 2024.

\bibitem{anwar} 	K. Anwar, {\it On the quantization of the electromagnetic field with magnetic monopoles}, arXiv:2509.17284.

\bibitem{vento} 	V. Vento, {\it Dark monopoles}, arXiv:2507.06697.

\bibitem{azevedo} D.O.R. Azevedo, T.S. Dias, E.D. Pereira, {\it The Euler-Heisenberg action for a $U(1)\times U(1)$ dyon quantum electrodynamics}, arXiv:2607.10906.

\bibitem{ATLAS:2023esy} G. Aad et al. (The ATLAS Collaboration), J. High Energy Phys. (JHEP) 11 (2023) 112.

\bibitem{becchi} C. Becchi, A. Rouet, R. Stora, Comm. Math. Phys. 42 (1975) 127; C. Becchi, A. Rouet, R. Stora, Ann. Phys. (N.Y.) 98 (1976) 287.

\bibitem{piguet-rouet} O. Piguet, A. Rouet, Phys. Rep. 76 (1981) 1.

\bibitem{faddeev1967feynman} L.D. Faddeev, V.N. Popov, Phys. Lett. B25 (1967) 29.

\bibitem{lautrup-nakanishi} N. Nakanishi, Progr. Theor. Phys. 35 (1966) 1111; Progr. Theor. Phys. 37 (1967) 618; B. Lautrup, Mat. Fys. Medd. Dan. Vid. Selsk 35 (11) (1967).

\bibitem{Low} J.H. Lowenstein, W. Zimmermann, Nucl. Phys. B86 (1975) 77; J.H. Lowenstein, Commun. Math. Phys. 47 (1976) 53; J.H. Lowenstein, {\it Renormalization Theory}, G. Velo, A.S. Wightman (Eds.), D. Reidel (Dordrecht-Holland), 1976.

\bibitem{del2016symanzik} O.M. Del Cima, D.H.T. Franco, O. Piguet, Nucl. Phys. B912 (2016) 51.

\bibitem{low-schroer} J.H. Lowenstein, B. Schroer, Phys. Rev. D6 (1972) 1553; J.H. Lowenstein, B. Schroer, Phys. Rev. D7 (1973) 1929.

\bibitem{Luders} G. L\"uders, B. Zumino, Phys. Rev. 106 (1957) 385; G. L\"uders, Ann. Phys. (N.Y) 2 (1957) 1.

\bibitem{Greenberg} O.W. Greenberg, Found. Phys. 36 (2006) 1535.

\bibitem{itzykson} C. Itzykson, J.-B. Zuber, {\it Quantum Field Theory}, Physics Series, McGraw-Hill, 1980.

\bibitem{qap} J.H. Lowenstein, Phys. Rev. D4 (1971) 2281; J.H. Lowenstein, Comm. Math. Phys. 24 (1971) 1; 
Y.M.P. Lam, Phys. Rev. D6 (1972) 2145; Y.M.P. Lam, Phys. Rev. D7 (1973) 2943; T.E. Clark, J.H. Lowenstein, Nucl. Phys. B113 (1976) 109.

\bibitem{piguet2} O. Piguet, S.P. Sorella, {\it Algebraic Renormalization}, Lecture Notes in Physics, m28, Springer-Verlag (Berlin-Heidelberg), 1995.

\bibitem{Wadojackiwpi} O.M. Del Cima, Phys. Lett. B720 (2013) 254.

\bibitem{chaichian} M. Chaichian, J. Fischer, Y.S. Vernov,  Nucl. Phys. B383 (1992) 151.

\bibitem{kugo} T. Kugo, I. Ojima, Progr. Theor. Phys. 60 (1978) 1869; T. Kugo, I. Ojima, Phys. Lett. B73 (1978) 459.

\bibitem{stora} R. Stora, {\it Renormalization Theory}, G. Velo, A.S. Wightman (Eds.), D. Reidel (Dordrecht-Holland), 1976.

\bibitem{kraus} E. Kraus, K. Sibold, Z. Phys. 68 (1995) 331.

\end{thebibliography}
\end{document}